# Evolution of a large online social network

Haibo Hu[*] and Xiaofan Wang

Complex Networks and Control Lab, Shanghai Jiao Tong University, Shanghai 200240, China

**Abstract:** Although recently there are extensive research on the collaborative networks and online communities, there is very limited knowledge about the actual evolution of the online social networks (OSN). In the letter, we study the structural evolution of a large online virtual community. We find that the scale growth of the OSN shows non-trivial S shape which may provide a proper exemplification for Bass diffusion model. We reveal that the evolutions of many network properties, such as density, clustering, heterogeneity and modularity, show non-monotone feature, and shrink phenomenon occurs for the path length and diameter of the network. Furthermore, the OSN underwent a transition from degree assortativity characteristic of collaborative networks to degree disassortativity characteristic of many OSNs. Our study has revealed the evolutionary pattern of interpersonal interactions in a specific population and provided a valuable platform for theoretical modeling and further analysis.

## 1. Introduction

Recently networks have constituted a fundamental framework for analyzing and modeling complex systems [1]. Social networks are typical examples of complex networks. A social network consists of all the people–friends, family, colleague and others–with whom one shares a social relationship, say friendship, commerce, or others. Traditional social network study can date back about half a century, focusing on interpersonal interactions in small groups due to the difficulty in obtaining large data sets [2]. The advent of modern database technology has greatly stimulated the statistical study of networks. Novel network structures of human societies have been revealed. The systematic research on the large scale available social network data has created a new subfield of network-sociology which integrates the theories of traditional social networks and modern complex networks [3].

As we know, now the WWW is undergoing a landmark revolution from the traditional Web 1.0 to Web 2.0 characterized by social collaborative technologies, such as Social Networking Site (SNS), blog, Wiki and folksonomy. In recent years, as a fast growing business, many SNSs of

---

[*] Corresponding author. Tel.: +86 21 34203083.

E-mail address: hbh@sjtu.edu.cn (H.-B. Hu).

differing scope and purpose have emerged in the Internet. The OSNs, constructed from the SNSs and embedded in Cyberspace, have attracted attentions of researchers from different disciplines, examples of which include *MySpace* [4], *Facebook* [5], *Pussokram* [6], etc. SNSs provide an online private space for individuals and tools for interacting with other people in the Internet. Thus the statistics and dynamics of these OSNs are of tremendous importance to researchers interested in human behaviors [3, 7].

In this letter, we will focus on an Internet community – *Wealink* (http://www.wealink.com), which is one of the largest OSNs in China at present. It is found that the scale growth of the OSN shows non-trivial S shape. At the same time the evolutions of network density, clustering, heterogeneity and modularity exhibit complex, non-monotone feature. We also observe shrink phenomenon for the path length and diameter of the network. Most interestingly it is found that the OSN has undergone a transition from degree assortativity characteristic of collaborative networks to degree disassortativity characteristic of many virtual communities. There can be different evolving mechanisms for the two kinds of social networks. Besides we also compare some observed quantities to the corresponding values from randomized networks with the same degree sequence as the original and reveal the significance of some specific network features.

## 2. Data sets

*Wealink* is a large SNS whose users are mostly professionals, typically businessmen and office clerks. Each registered user of the SNS has a profile, including his/her list of friends. If we view the users as nodes $V$ and friend relationships as edges $E$, an undirected friendship network $G(V,E)$ can be constructed from *Wealink*. For privacy reasons, the data, logged from 0:00:00 h on 11 May 2005 (the inception day for the Internet community) to 15:23:42 h on 22 Aug 2007, include only each user's ID and list of friends, and the establishment time for each friend relationship. The online community is a dynamical evolving one with the new users joining in the community and new connections established between users.

## 3. Structural evolution

We extract 27 snapshots of *Wealink* with an interval of one month from June 11[th], 2005 to August 11[th], 2007 and investigate the evolution of the network. Fig. 1 shows the growth of the numbers of nodes and edges, and the variation of network density over time. The density of a network is defined as the ratio of the number of real edges $E$ to the number of total possible edges $N(N-1)/2$ ($N$ is the number of nodes). In a recent work, Leskovec *et al.* observed that certain citation graphs, the Internet at the Autonomous System level, and email networks became denser over time, with the number of edges growing super-linearly in the number of nodes [8]. However, we find that the density of the OSN as a function of time is non-monotone. There are three obviously marked stages: an initial upward trend leading to a peak, followed by a decline, and the final gradual steady increase, which also is observed in *Flickr* and *Yahoo! 360* studied by Kumar *et al* [9]. A possible reason is that right after the establishment of the OSN, there was an initial excitement among a few enthusiasts who joined the network and frantically invited many of their friends to join; this gives rise to the first stage that culminates in a peak. The second stage corresponds to a natural dying-out of this euphoria and this leads to the decline. Finally the network appears to arrive at equilibrium and its density seems to converge to constant. Some network models based on preferential attachment cannot reproduce this phenomenon.

It is interesting that the growth pattern for the network scale, including the numbers of nodes and edges, shows S shape. It is well known that S-shaped curve is the heart of many diffusion processes and is characteristic of a chain reaction, in which the number of people who adopt a new behavior follows a logistic-like function [10]: a slow growth in the initial stage, rapid growth for critical mass time, and a rapid flattening of the curve beyond this point. We may view membership and friendship in the OSN as a 'behavior' that spreads through the Chinese professionals. According to Bass diffusion model [11] and as shown in Fig. 1, the growth of network scale $S(T)$ can be fitted by the logistic function $S(T) = \frac{m}{1+e^{-(T-a)s}}$. S-shaped growth of network scale was also observed by Chun *et al*. in *Cyworld* [12]. For large $T$, the increment of the number of nodes/edges, i.e. $dS/dT$, is exponentially decaying. A recent research has shown that the weekly overall grosses income of movies, reflecting the diffusion of a movie in population, indeed decay exponentially with time [13].

Fig. 2 shows the numbers of new nodes and edges in the $T$ th month for the whole network. Most users and links appeared in the network over a short period of time, i.e. between the 11[th] and 17[th] month. After the 17[th] month, every month there were still some new users joining in the network and new links created in the network, although the numbers of them are small. It is clear that each new link can be created between two old users (Old-Old), two new users (New-New) or an old user and a new one (New-Old). According to the types of users the newly established edges can be divided into three classes. Fig. 3 shows the proportion of three kinds of edges created in the $T$ th month. We find that most links were created between two old users or an old user and a new one, even after the 17[th] month. The results indicate that the evolving network has reached a state of equilibrium.

We evaluate the shortest path lengths $l_{ij}$ ( $N \geq i > j \geq 1$ ) of every node pair within the social network, and obtain the average path length $l = \sum l_{ij} / C_N^2$. The diameter of the network is $D = \max l_{ij}$. Fig. 4 shows the evolution of $l$ and $D$ for the largest connected subgraph over time compared to network density $d$. The growth of $d$ may provide partial explanation for the evolution pattern of $l$ and $D$ [9]. The evolutions of both $l$ and $D$ also show three different stages: in the first stage, they are almost constants; in the next stage, when the network density declines, the $l$ and $D$ grow till they reach peaks; in the third stage, when the $d$ starts increasing slowly, accordingly the $l$ and $D$ start minishing slowly. However it is clear that there is no obvious evidence that these two metrics are analytically connected to each other. Recently the phenomenon of shrinking diameter was also observed by Kumar *et al.* [9] in *Flickr* and *Yahoo! 360* and by Leskovec *et al.* [8] in citation graphs and the Internet at the Autonomous System level. A network model based on forest fire spreading process can reproduce the feature of shrinking diameter [8].

Social networks are believed to have many connected triads, i.e. a high degree of transitivity. That is, if person *A* knows *B* and *C*, then it is quite likely that *B* and *C* are acquainted. A quantity to measure the strength of connections within $i$'s neighborhood is the local clustering coefficient $c_i = e(\Gamma_i)/C^2_{k_i}$, where $e(\Gamma_i)$ is the number of real edges within $i$'s neighborhood $\Gamma_i$ consisting of $k_i$ nodes, and $C^2_{k_i}$ is the total number of all possible connections in $\Gamma_i$ [14]. The clustering coefficient $C$ of the entire network is defined as the mean value of $c_i$.

Fig. 5(a) shows the time evolution of clustering coefficients $C$ for the largest connected subgraph. It shows strong positive correlation with density $d$ in the network. And almost at the same time $C$ and $d$ arrive at their equilibrium values. High network density favors clustering to some extent. However there is also no distinct evidence that the two metrics are analytically correlative with each other. For a random network, although its density may be large, its clustering coefficient still may be less than that of a low-density network with, say, strong community feature. Fig. 5(b) shows the comparison of $C$ between actual networks and reshuffled ones obtained by random degree-preserving rewiring of the original networks [15]. It is evident that the $C$ of actual networks is significantly larger than that of the randomized and it is similar in growth trend to that of randomized ones.

The degree distribution for the largest connected subgraph at the largest time can be fitted by a power law $P(k) \sim k^{-\gamma}$ with a degree exponent $\gamma \approx 1.74$. Compared to exponential networks, power law networks are thought to be highly heterogeneous. The heterogeneity of node degrees of a network can be quantified by heterogeneity index $H$, which can be computed as $H = \sum_{i=1}^{N}\sum_{j=1}^{N}|k_i - k_j|/2N^2\langle k \rangle$, where $k_i$ ($1 \leq i \leq N$) is node degree, $\langle k \rangle$ is the mean degree, and $0 \leq H < 1$ [16]. The $H$ of exponential networks has maximum 0.5. The variation of $H$ for the largest connected subgraph is shown in Fig. 6. It displays non-monotone feature and finally converges to a constant.

Fig. 7 shows the evolution of the degree distribution of the largest connected subgraphs of the network. The degree distribution gradually converges to a power law with degree exponent $\gamma \approx 1.74$. In *Wealink*, users can create 'discussing groups' with specific topics. The users who are interested in some topic can join in the corresponding group and create friend relations with the users in the group. Each discussing group comprises 30 users at most. As shown in Fig. 7, the function of *Wealink* leads to the emergence of peak values of degree distribution at degrees of about integral multiples of 30 (30, 60, 90, etc.).

Many social networks show distinct feature of community structure, the gathering of nodes into groups such that there is a higher density of edges within groups than between them. We analyze the community structure of the largest connected subgraph of the OSN with the hierarchical agglomeration algorithm proposed by Clauset et al [17]. Modularity $Q$ is a measure of the quality of a particular division of a network into communities, which measures the fraction of the edges in the network that connect nodes of the same type (i.e., withincommunity edges) minus the expected value of the same quantity in a network with the same community divisions but random connections between the nodes. $Q$ is defined as follows [18]: Consider a specific division of a network into communities, and suppose $e_{ij}$ is the fraction of all edges in the network that link nodes in community $i$ to nodes in community $j$, then $\sum_i e_{ii}$ gives the fraction of edges that join nodes in the same community $i$. Define $a_i = \sum_j e_{ij}$ represents the fraction of edges that connect to nodes in community $i$, then $Q = \sum_i (e_{ii} - a_i^2)$. A $Q$ value above about 0.3 is a good indicator of significant community structure in a network.

Fig. 8 shows the evolution of $Q$ for the largest connected subgraph in comparison with the same metric of randomized networks. On the whole $Q$ shows increasing trend indicating the modularity of the network is more and more prominent, and finally the network gets to equilibrium with almost changeless $Q$. In the online professional social network, the tendency of establishing interpersonal relations among the individuals with the same or similar occupations and the existence of discussing groups with specific topics, can lead to the emergence of communities with densely interconnected individuals and the high modularity of the network. The *Wealink* has significantly higher $Q$ than those of randomized networks and the $Q$ of actual networks is similar in growth trend to those of randomized ones. We also find that the randomized networks still possess large $Q$, which may result from the structural constraint of degree sequence of original networks.

The correlation can exist between the degrees of adjacent nodes, which is often characterized by the assortativity $r$ and defined as the Pearson correlation coefficient of the degrees of either nodes connected by a link: $r = (\langle ij \rangle - \langle i \rangle \langle j \rangle)/(\langle i^2 \rangle - \langle i \rangle^2)$, where $i$ and $j$ are the remaining degrees at the two ends of an edge and the $\langle \cdot \rangle$ notation represents the average over all links [19]. If a network's assorativity is negative, a hub tends to be connected to non-hubs, and vice versa. When $r > 0$, we call the network to have an assortative mixing pattern, and when $r < 0$, disassortative mixing. The conventional wisdom is that social networks exhibit an assortative mixing pattern, whereas biological and technological networks show a disassortative mixing pattern [20]. However the recent research on the OSNs modifies the wide-spread belief and many OSNs show a disassortative or neutral mixing trait, which is displayed in Table 1.

The origins of obvious degree assoratity for collaborative networks are miscellaneous. From the perspective of sociology and psychology, in real life everyone would like to interact with elites in the society; however the elites would rather communicate with the people with the same social status as theirs, which may lead to the assortative mixing pattern in the collaborative networks. For professional collaborations, such as scientific, actor, and business collaborations, the already big names preferably collaborate with other big names for success, reputation and influence. As indicated by Holme *et al*. [6], assortative mixing may be significant only to interaction in competitive areas. Besides it appears that some of the degree correlation in collaborative networks could have real organizational origins. Generally the networks of collaborations between academics, actors, and businessmen are affiliation networks, in which people are connected together by membership of common groups (authors of a paper, actors in a film, researchers in a lab, office clerks in a company, etc.) [28]. OSNs differ from the collaborative networks in these regards. They break the invisible boundary among different estates in a society. In the virtual world elites will not refuse connections from anyone because they know that more connections show others they are elites. And unlike in real life, these links are not costly. Relationships in the real world have to be maintained and this requires continual effort. The basic difference could be the deciding difference between virtual and real world. Understanding the process, the generative mechanism, will supply a substantial comprehension of the formation and evolution of online virtual communities.

Generally it is thought that real-world networks always belong to the same type over time, either assortative or disassortative. However as shown in Fig. 9, the *Wealink* underwent a transition from the initial assortativity characteristic of collaborative networks to subsequent disassortativity characteristic of many OSNs. To the best of our knowledge, this is the first real-world network observed which possesses the intriguing feature. A reasonable conjecture is that often the friendship relations in the beginning OSN are based on real-life interpersonal relations, which means that *Wealink* users were linking to the other users who are their friends in the real world. In this case the OSN directly inherits the assortative structure of the underlying real-life interpersonal network. However at the later stage many online users of low degrees may preferentially establish connections with the network elites of high degrees, resulting in the disassortative mixing.

Fig. 9 also shows the comparison of $r$ between actual networks and randomized ones. The randomized networks show disassortative or nearly neutral mixing feature, and all $r$'s are less than 0 and no transition appears. The comparison shows that the *Wealink* is strongly degree assortative at the beginning stage and disassortative at the later stage, suggesting that individuals indeed draw their partners from the users with degrees similar (beginning) or dissimilar (later) to theirs far more often than one would expect on the basis of pure chance.

## 4. Summary and discussion

In the letter, we have studied the structural evolution of a large OSN. The selection of friends performed by registered users is driven by their personal occupational background. We have found that the scale growth of the OSN shows S shape, which may provide an exemplification for Bass diffusion model. The evolutions of network density, clustering, heterogeneity and modularity show non-monotone feature, and shrink phenomena for the path length and diameter of the network occur. Especially we have shown that the OSN underwent a transition from degree assortativity to disassortativity. The OSN has arrived at its equilibrium, where its statistical quantities have converged to their constant limits.

A social network might be divided up according to the location, affiliation, occupation, and so forth, of its members. It is thought that clustering and assortativity in networks arise because the vertices are divided into communities [20]; however from Figs. 5, 8 and 9 we find that both clustering coefficient and degree assortativity are negatively correlative with modularity. Recent study also showed that the heterogeneity of networks could be negatively correlative with assortativity coefficient [29]; the comparison between Fig. 6 and Fig. 9 may provide a proper counterexample to the prevailing view. It should be quite cautious to claim that there exists specific correlation between network metrics, and obviously under different conditions there may be quite different conclusions. Different network properties, such as modularity, clustering, assortativity, heterogeneity, synchronizability, etc., may constrain each other, or not be independent [30]. However for a network with some fixed metric, say, degree exponent $\gamma$, by random rewiring, we also can obtain the tunable space for another network metric, say assortativity coefficient $r$. And within the $\gamma$-$r$ space, under disparate conditions, all the three cases, positive correlation, negative correlation and no correlation between $\gamma$ and $r$, can occur.

Until now it is still not completely known how the various topological properties are related, or if it is possible to single out a family of metrics defining all others, which constitutes an open field of study [31].

As a rapidly developing field in interdisciplinary research, virtual community has attracted scholars of different backgrounds, mostly physicists and computer scientists. However the main body in the virtual world is still persons in real world, thus as pointed out by Tim Berners-Lee - the "father of the World Wide Web", understanding the web community may also require insights from sociology and psychology every bit as much as from physics and computer science [32].

**Acknowledgments**

We thank Wealink Co. for providing the network data. This work was partly supported by the NSF of PRC under Grant No. 60674045.

**Figure Captions**

Fig. 1. Evolution of the numbers of nodes $N$ and edges $E$, and network density $d$ for the whole network (a) and the largest connected subgraph (b) of *Wealink* from June 11[th] 2005 to August 11[th] 2007 with an interval of one month. Solid lines are the fitted curves by the logistic function. In (a) least squares fitting gives $m = 2.244 \times 10^5 \pm 2.813 \times 10^3$, $a=12.504 \pm 0.092$ and $s=1.524 \pm 0.185$ for $N(T)$ and $m=2.721 \times 10^5 \pm 2.604 \times 10^3$, $a=12.704 \pm 0.073$ and $s=1.358 \pm 0.118$ for $E(T)$. In (b) least squares fitting gives $m = 1.943 \times 10^5 \pm 2.071 \times 10^3$, $a=12.600 \pm 0.078$ and $s=1.527 \pm 0.157$ for $N(T)$ and $m=2.443 \times 10^5 \pm 2.265 \times 10^3$, $a=12.734 \pm 0.072$ and $s=1.306 \pm 0.107$ for $E(T)$.

Fig. 2. The number of new nodes/edges at $T$, i.e. the increment of the number of nodes/edges at $T$.

Fig. 3. Evolution of the proportion of three kinds of edges.

Fig. 4. Time evolution of average path length $l$ and diameter $D$ for the largest connected subgraph of *Wealink* in comparison with network density $d$.

Fig. 5. (a) Variation of clustering coefficient $C$ for the largest connected subgraph of *Wealink* over time in comparison with network density $d$. (b) The left scale is for clustering coefficient of largest connected subgraph of *Wealink* and its randomized version and the right scale the ratio of $C$ of actual networks to that of randomized ones.

Fig. 6. Variation of heterogeneity index $H$ for the largest connected subgraph of *Wealink* over time.

Fig. 7. Evolution of the degree distribution of *Wealink*.

Fig. 8. Evolution of modularity $Q$ for the actual networks and corresponding randomized ones.

Fig. 9. Evolution of degree assortativity $r$ for the *Wealink* and its randomized version.

Table 1. Degree assortativity coefficients of OSNs and collaborative networks. The percentage in parenthesis indicates sampling ratio.

| Type | Network | $N$ | $r$ | References |
|---|---|---|---|---|
| Online social network | Cyworld | 12, 048, 186 | -0.13 | [4] |
| | nioki | 50, 259 | -0.13 | [6] |
| | All contacts in pussokram | 29, 341 | -0.05 | [6] |
| | Messages in pussokram | 21, 545 | -0.06 | [6] |
| | Guest book in pussokram | 20, 691 | -0.07 | [6] |
| | Friends in pussokram | 14, 278 | -0.04 | [6] |
| | Flirts in pussokram | 8, 186 | -0.12 | [6] |
| | MySpace | 100, 000 (~0.08%) | 0.02 | [4] |
| | orkut | 100, 000 (~0.30%) | 0.31 | [4] |
| | Xiaonei | 396, 836 | -0.0036 | [21] |
| | Gnutella P2P(SN 6) | 191, 679 | −0.109 | [22] |
| | Flickr | 1, 846, 198 (26.9%) | 0.202 | [23] |
| | LiveJournal | 5, 284, 457 (95.4%) | 0.179 | [23] |
| | YouTube | 1, 157, 827 | -0.033 | [23] |
| | mixi | 360, 802 | 0.1215 | [24] |
| Collaborative network | ArXiv coauthorship | 52, 909 | 0.36 | [25] |
| | Cond-mat coauthorship | 16, 264 | 0.18 | [25] |
| | Mathematics coauthorship | 253, 339 | 0.12 | [19] |
| | Neuroscience coauthorship | 205, 202 | 0.60 | [26] |
| | Biology coauthorship | 1, 520, 251 | 0.13 | [25] |
| | Film actor collaboration | 449, 913 | 0.21 | [19] |
| | TV series actor collaboration | 79, 663 | 0.53 | [27] |
| | Company directors | 7, 673 | 0.28 | [19] |

Fig. 1

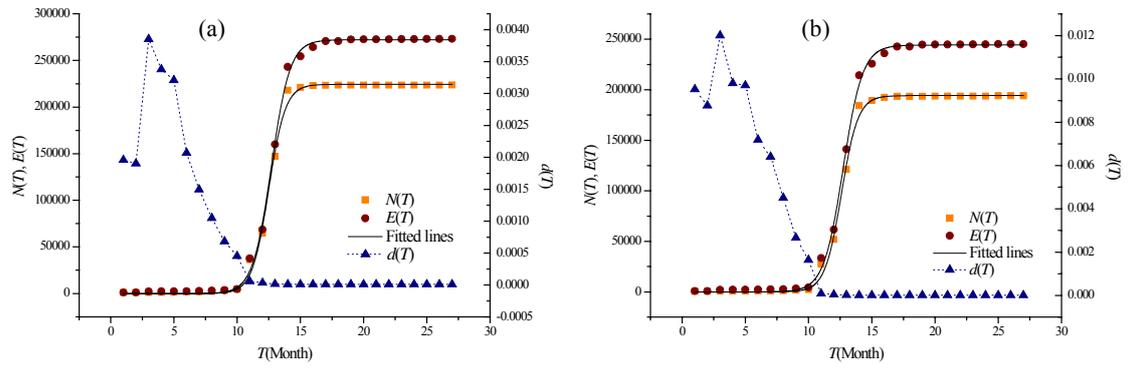

Fig. 2

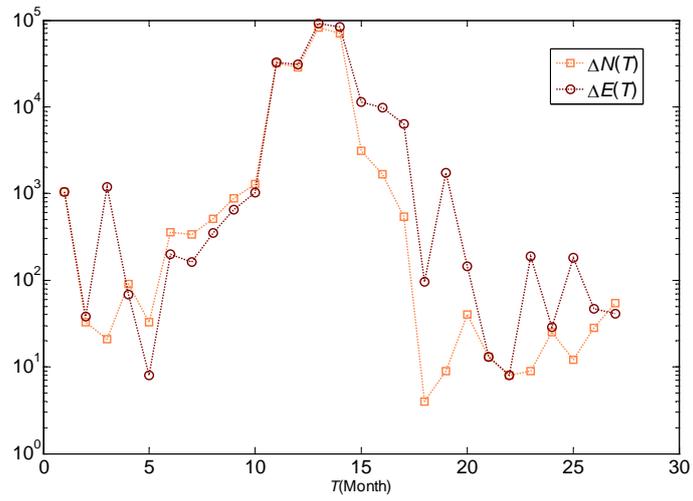

Fig. 3

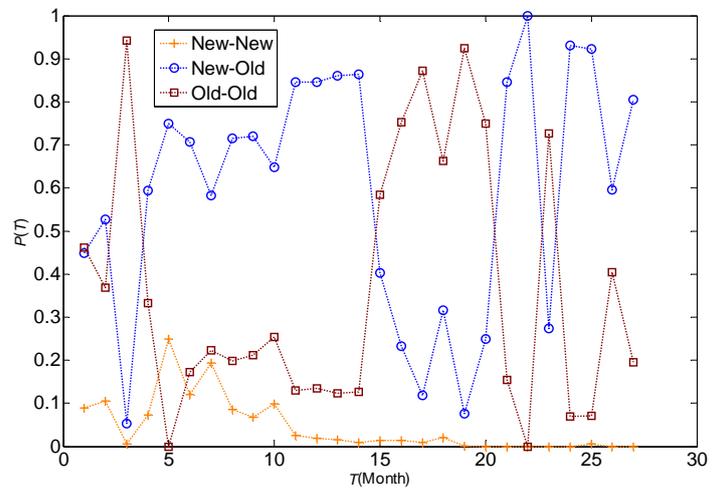

Fig. 4

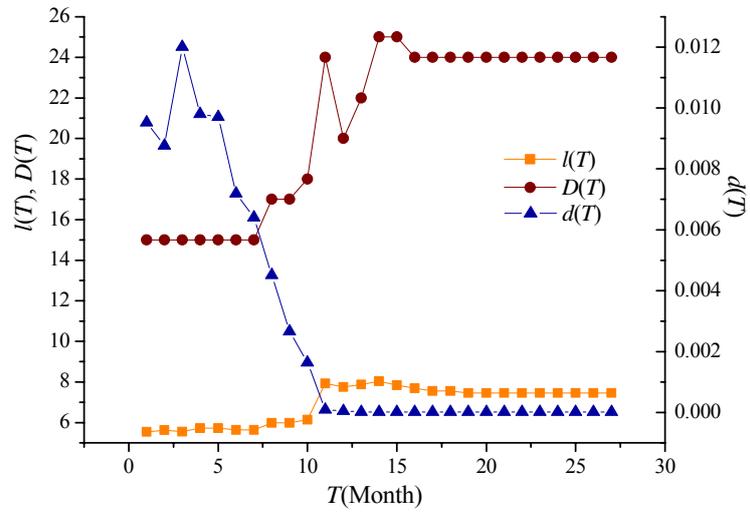

Fig. 5

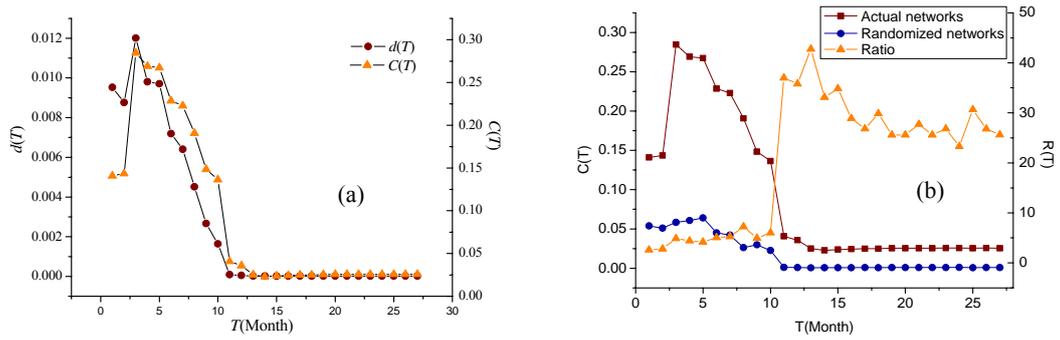

Fig. 6

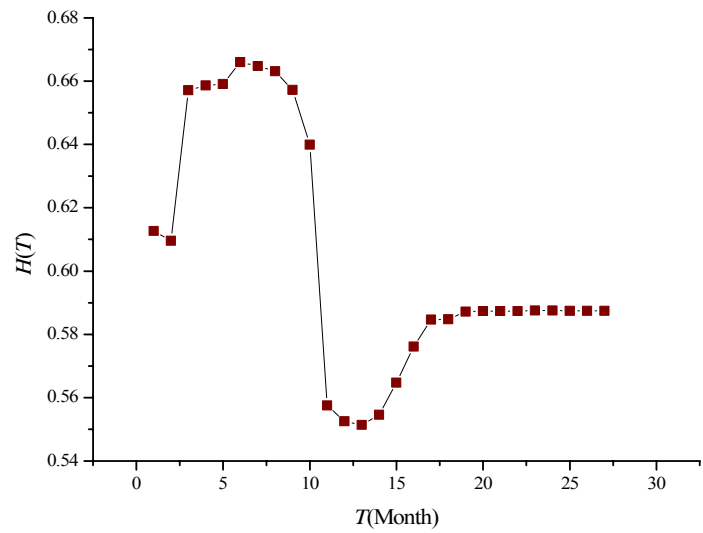

Fig. 7

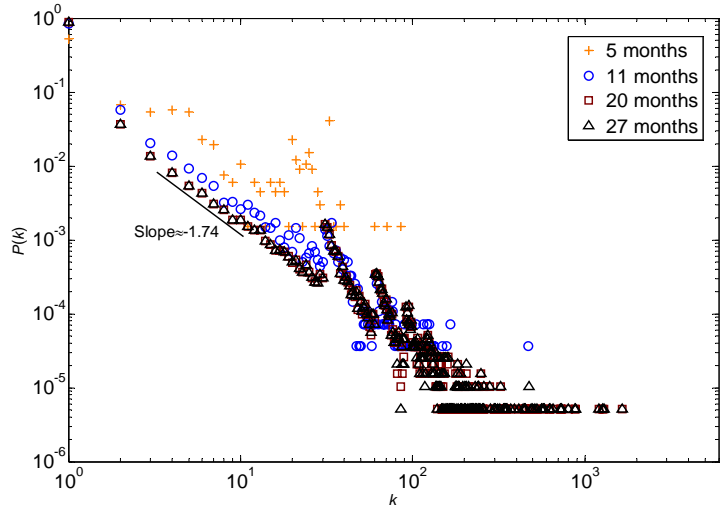

Fig. 8

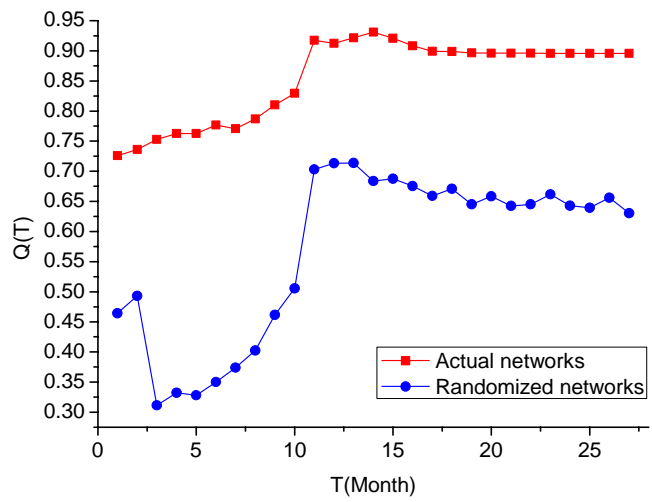

Fig. 9

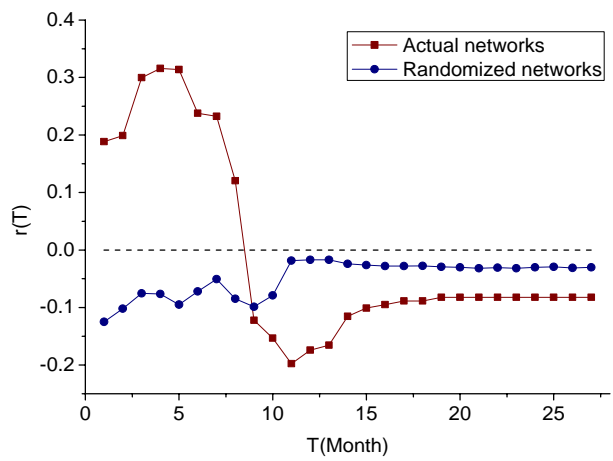